\documentclass[conference]{IEEEtran}
\IEEEoverridecommandlockouts

\usepackage[T1]{fontenc}

\ifCLASSOPTIONcompsoc
  \usepackage[nocompress]{cite}
\else
  \usepackage{cite}
\fi

\ifCLASSINFOpdf
\else
\fi

\usepackage{amsmath}
\usepackage{cite}
\usepackage{times}

\usepackage{graphicx}
\graphicspath{{figs/}}
\usepackage{subfigure}
\usepackage{color}
\usepackage{ifpdf}
\usepackage{epsfig}
\usepackage{latexsym}
\usepackage{amsfonts}
\usepackage{amssymb}
\usepackage{paralist}
\usepackage{comment}
\usepackage{xspace}
\usepackage{mathrsfs}
\usepackage{epsf}
\usepackage{amssymb}
\usepackage{setspace}
\usepackage{color}
\usepackage{caption}

\usepackage{algorithm}
\usepackage{algorithmic}

\usepackage{amssymb}
\usepackage{url,epsfig,array}
\usepackage{leftidx}
\usepackage{epstopdf}
\usepackage{diagbox}
\usepackage{makecell}
\usepackage{multirow}
\usepackage{booktabs}
\usepackage{enumerate}
\usepackage{pifont}
\usepackage{threeparttable}

\usepackage{amsthm}
\usepackage{balance}
\usepackage{bbding}

\allowdisplaybreaks[4]

\usepackage{array}
\newcommand{\PreserveBackslash}[1]{\let\temp=\\#1\let\\=\temp}
\newcolumntype{C}[1]{>{\PreserveBackslash\centering}p{#1}}
\newcolumntype{R}[1]{>{\PreserveBackslash\raggedleft}p{#1}}
\newcolumntype{L}[1]{>{\PreserveBackslash\raggedright}p{#1}}

\def\ie{\textit{i.e.}\xspace}
\def\etal{\textit{et al.}\xspace}

\def\eg{\textit{e.g.}\xspace}

\def\st{\xspace\textbf{s.t.}\xspace}

\newtheorem{theorem}{Theorem}

\newtheorem{corollary}{Corollary}

\newtheorem{lemma}{Lemma}

\newtheorem{assumption}{Assumption}

\newcommand{\tabincell}[2]{\begin{tabular}{@{}#1@{}}#2\end{tabular}}

\renewcommand{\algorithmicrequire}{\textbf{Input:}}
\renewcommand{\algorithmicensure}{\textbf{Output:}}

\newcommand{\argmin}{\operatornamewithlimits{argmin}}

\newcommand{\inst}[1]{\textsuperscript{#1}}

\begin{document}


\title{Air-FedGA: A Grouping Asynchronous Federated Learning Mechanism Exploiting Over-the-air Computation}

\author{Qianpiao Ma\inst{1}, Junlong Zhou\inst{1}\Envelope, Xiangpeng Hou\inst{1}, Jianchun Liu\inst{2}, Hongli Xu\inst{2}, Jianeng Miao\inst{2}, Qingmin Jia\inst{3}\\
\inst{1}Nanjing University of Science and Technology, Nanjing, China, \{maqianpiao,jlzhou,xphou\}@njust.edu.cn \\
\inst{2}University of Science and Technology of China, Hefei, China,\\
 \{jcliu17,xuhongli\}@ustc.edu.cn, canon@mail.ustc.edu.cn \\
\inst{3}Purple Mountain Laboratories, Nanjing, China, jiaqingmin@pmlabs.com.cn	

\IEEEcompsocitemizethanks{\IEEEcompsocthanksitem This paper has been accepted by IPDPS 2025. \protect
}

}

\maketitle

\begin{abstract}
Federated learning (FL) is a new paradigm to train AI models over distributed edge devices (\ie, workers) using their local data, while confronting various challenges including communication resource constraints, edge heterogeneity and data Non-IID. Over-the-air computation (AirComp) is a promising technique to achieve efficient utilization of communication resource for model aggregation by leveraging the superposition property of a wireless multiple access channel (MAC). However, AirComp requires strict synchronization among edge devices, which is hard to achieve in heterogeneous scenarios. In this paper, we propose an AirComp-based grouping asynchronous federated learning mechanism (Air-FedGA), which combines the advantages of AirComp and asynchronous FL to address the communication and heterogeneity challenges. Specifically, Air-FedGA organizes workers into groups and performs over-the-air aggregation within each group, while groups asynchronously communicate with the parameter server to update the global model. In this way, Air-FedGA accelerates the FL model training by over-the-air aggregation, while relaxing the synchronization requirement of this aggregation technology.
We theoretically prove the convergence of Air-FedGA. We formulate a training time minimization problem for Air-FedGA and propose the power control and worker grouping algorithm to solve it, which jointly optimizes the power scaling factors at edge devices, the denoising factors at the parameter server, as well as the worker grouping strategy. We conduct experiments on classical models and datasets, and the results demonstrate that our proposed mechanism and algorithm can speed up FL model training by 29.9\%-71.6\% compared with the state-of-the-art solutions.

\end{abstract}


\section{Introduction}\label{sec:intro}

The rapid growth of the Internet of Things (IoT) has led to the generation of massive amounts of data from edge devices, such as sensors, mobile phones, and base stations \cite{zhu2018towards}. Effectively utilizing this data is crucial for enhancing the quality of service (QoS) on various applications, such as interactive online gaming, face recognition, 3D modeling, VR/AR and vehicle networking systems. Recently, artificial intelligence (AI) algorithms have deployed from the centralized cloud to the distributed network edge, which is known as edge AI \cite{wang2019edge}, enabling efficient data processing locally. This shift has facilitated the adoption of federated learning (FL), a technique that trains AI models over edge nodes while preserving privacy by utilizing data locally.

FL has gained significant attention since its introduction in 2016 \cite{konevcny2016federated}.
A typical FL system usually consists of a large number of edge devices (\ie, workers) for local model training, and a centralized parameter server (PS) for each round of local model aggregation \cite{li2014scaling}.
Each worker trains model over its local dataset and sends the trained local model to the PS. The PS aggregates these local models into a global model and broadcasts it back to the workers. This procedure continues for multiple rounds until the global model converges.

However, when deployed at the network edge, FL faces several challenges: \textbf{1) Limited communication resource}: Traditional orthogonal multiple access (OMA) schemes (TDMA \cite{tran2019federated,mo2021energy,luo2021cost}, OFDMA \cite{chen2020convergence}\cite{yang2020energy}) are commonly used for FL model aggregations. However, due to limited communication resources, the transmission delay increases linearly with the number of workers when deploying OMA schemes\cite{zhu2020broadband}, resulting in unsatisfactory scalability in large-scale FL scenarios. \textbf{2) Edge heterogeneity}: The CPU capacities, data sizes, and network connections are usually heterogeneous at network edge. As a result, the required time to perform local updating and receive/upload models may vary significantly. \textbf{3) Data Non-IID}: A device's local data are often not sampled drawn uniformly from the overall distribution. In other words, the local data on edge devices are typically non-independent and non-identically distributed (Non-IID) \cite{mcmahan2017communication-efficient}.

Traditional OMA schemes aggregate models on the premise that all model vectors can be reliably transmitted. However, the parameter server actually just needs to obtain the weighted average of the local model vectors, rather than ensuring reliable transmission of each individual vector \cite{fan2021joint}. To this end, \emph{over-the-air computation} (AirComp) \cite{nazer2007computation}, a new analog non-orthogonal multiple access (NOMA) technique, is motivated to break through the limitation of communication resources, and reduce transmission delay for large-scale FL \cite{cao2021optimized,zhu2020one,yang2020federated,amiri2020machine,cao2022transmission,amiri2020federated}. AirComp-based aggregation is achieved by synchronizing workers to transmit their local model vectors concurrently, leveraging the superposition property of wireless multiple access channels (MAC) to sum these vectors over-the-air \cite{cao2021optimized}.

\begin{table*}[htb]\centering
{
{
\caption{Performance comparison for FL mechanism.}\label{tbl:compare}
\begin{tabular}{c|c|c|c|c}
\hline
FL Mechanism & \makecell{Communication\\ Comsuption} & \makecell{Handing Edge \\Heterogeneity} & \makecell{Handing Non-IID} & Scalability\\
\hline
Synchronous\cite{konevcny2016federated,tran2019federated,mo2021energy,luo2021cost,chen2020convergence,yang2020energy,mcmahan2017communication-efficient} & Medium & Poor & Medium & Poor \\
AirComp+Synchronous\cite{zhu2020broadband,cao2021optimized,zhu2020one,yang2020federated,amiri2020machine,cao2022transmission,amiri2020federated} & Low & Poor & Medium & Poor \\
Asynchronous\cite{wu2020safa,xie2019asynchronous,chen2020asynchronous,zheng2017asynchronous,zhu2022client,ma2021fedsa,chai2020tifl,liu2021adaptive} & High & Good & Poor & Good \\
AirComp+Asynchronous (\textbf{Air-FedGA}) & Low & Good & Good & Good \\
\hline
\end{tabular}
}\vspace{-2mm}
}
\end{table*}

\begin{figure}[t]\centering
\includegraphics[width=0.49\textwidth]{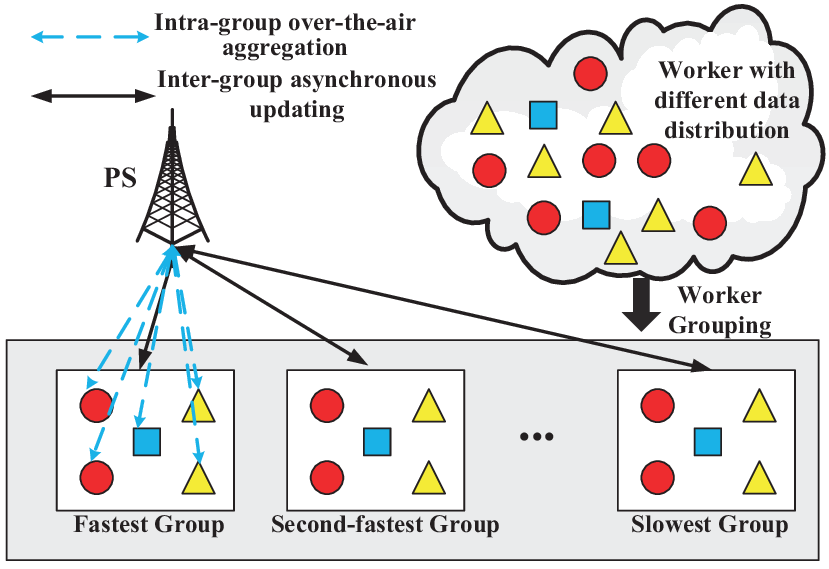}
\caption{The Architecture of Air-FedGA.} \label{fig:architecture}
\vspace{-2mm}
\end{figure}

The implementation of AirComp has a bifacial impact on FL. On one hand, due to the efficient spectrum utilization of AirComp-based aggregation, it is expected to significantly reduce the transmission delay compared to the OMA-based aggregation which decouples communication and computation.
On the other hand, strict synchronization among all heterogeneous workers is required for successful over-the-air aggregation \cite{zhu2020broadband}. However, due to edge heterogeneity, the completion time of workers varies significantly \cite{wu2020safa}. As a result, the parameter server has to wait for the slowest worker to complete its local training before the over-the-air aggregation, while other workers are idle at this time. It is known as the \emph{straggler problem} \cite{xie2019asynchronous}, leading long single-round duration and decelerating FL. Conventionally, asynchronous FL mechanisms are deployed to tackle the straggler problem \cite{wu2020safa,xie2019asynchronous,chen2020asynchronous,zheng2017asynchronous,zhu2022client,ma2021fedsa,chai2020tifl,liu2021adaptive}. In this case, the parameter server updates the global model with each local model as soon as it arrives, without waiting for others. However, a fully asynchronous mechanism cannot be directly applied to AirComp-based aggregation, since the implementation of AirComp is based on the concurrent transmission of multiple devices.

To address the difficulty of utilizing asynchronous mechanism for over-the-air FL, this paper propose an AirComp-based grouping asynchronous federated learning mechanism (Air-FedGA) over a noisy fading MAC. As shown in Fig. \ref{fig:architecture}, the workers in the FL system are organized into groups. The workers within a group perform over-the-air aggregation simultaneously, whereas each global updates is performed asynchronously among groups. Air-FedGA relaxes the requirement on synchronization of AirComp-based aggregation to accelerate FL, in which a worker only needs to wait for the worker within the same group to complete local training, instead of waiting for all workers. The comparison of different FL mechanisms are summarized in Table \ref{tbl:compare}.

The main contributions of this paper are as follows:
\begin{itemize}
    \item We design Air-FedGA, a group asynchronous FL mechanism via over-the-air computation, to address the communication constraint and edge heterogeneity. We analyze the convergence of Air-FedGA, and explore the quantitative relationship between the convergence bound and several factors, \eg, data distribution, the transmission power scaling factors and the denoising factors.
    \item We formulate a training time minimization problem for Air-FedGA. To solve this problem, we first jointly optimizes the power scaling factors at workers and the denoising factors at parameter server. Then we propose a worker grouping algorithm based on both on communication and data distribution. This approach ensures that, within each group, the local training times of workers are similar, while also striving to make the inter-group data distribution as close to IID as possible to mitigate the effects of Non-IID data.
    \item Experimental results on the classical models and datasets show that, compared with the state-of-the-art solutions, our proposed mechanism and algorithm can greatly accelerated FL model training by 29.9\%-71.6\%.
\end{itemize}

The rest of this paper is organized as follows. Section \ref{sec:related} reviews the related works. Section \ref{sec:prelim} introduces the grouping asynchronous federated learning via AirComp, and Section \ref{sec:convergence} provides its convergence analysis. The worker grouping algorithm is proposed in Section \ref{sec:algorithm}. The experimental results are shown in Section \ref{sec:evaluation}. Section \ref{sec:conclusion} concludes this paper.

\section{Related Works}\label{sec:related}

\subsection{Asynchronous Federated Learning}\label{subsec:airfl}

Asynchronous federated learning (FL) addresses the straggler problem by aggregating a local model with the global model as soon as the parameter server receives it, without waiting for all workers to finish their local training. However, this comes at the expense of using out-of-date models, inevitably incurring \textit{staleness} concern \cite{xie2019asynchronous}.

Many researches have focused on addressing edge heterogeneity with asynchronous scheme while mitigating the adverse effects of staleness. For example, Xie \etal \cite{xie2019asynchronous} assign smaller aggregation weights to stale models to lessen their impact on training. Wu \etal \cite{wu2020safa} propose a simple approach to handle staleness, where the parameter server discards too stale models during training process. Chen \etal \cite{chen2020asynchronous} deploy dynamic learning rates for workers according to the frequency of their participating in global updating, which can also alleviate the staleness concern. Zheng \etal \cite{zheng2017asynchronous} and Zhu \etal \cite{zhu2022client} compensate delayed gradients based on approximate Taylor expansion. Ma \etal \cite{ma2021fedsa} introduce a semi-asynchronous FL mechanism that involves multiple workers in each global updating to ensure that the model is not too stale. Chai \etal \cite{chai2020tifl} organize workers into groups according to their communication time with the parameter server, and perform global updating asynchronously among groups.

However, these works do not explicitly handle data Non-IID, which can amplify the negative effects of staleness and lead to gradient divergence \cite{liao2023decentralized}. In contrast, we propose a novel worker grouping algorithm that considers both the \textit{communication time} and the \textit{data distribution} of workers, and assigns them to different groups accordingly. Therefore, our algorithm can reduce the communication overhead and improve the convergence of FL under data Non-IID.

\subsection{Federated Learning via Over-the-air Computation}\label{subsec:airfl}

The first work that introduced over-the-air computation-based FL aggregation was by Zhu \etal \cite{zhu2020broadband}, who leverage the broadband analog aggregation to achieve low-latency model aggregation.
They further expand their work by using one-bit quantization at workers, followed by modulation and majority-voting-based decoding at the parameter server, to reduce the communication overhead \cite{zhu2020one}.
Yang \etal \cite{yang2020federated} focus on the trade-off between communication and learning, and propose a method that maximizes the number of devices while minimizing the mean squared error (MSE) of gradient error.
Another challenge in this approach is the bandwidth consumption. Mohammadi \etal \cite{amiri2020machine}\cite{amiri2020federated} exploit the sparsity of the model update vector and project it into a low-dimensional space using random matrices. Their methods significantly reduce the bandwidth requirement, while preserving the accuracy of the model aggregation. Power control is another important factor that affects the performance of FL via AirComp. Zhang \etal \cite{zhang2021gradient} formulate the power control problem as an optimization problem that minimizes the MSE of gradients, subject to average power constraints at each worker. Similarly, Cao \etal \cite{cao2022transmission} conduct an analysis of the convergence of over-the-air computation FL under various power control policies to optimize transmit power.

However, a common limitation in the aforementioned works is the requirement for all workers to concurrently transfer their local models for over-the-air aggregation, leading to the critical straggler problem. In response, our proposed group asynchronous mechanism allows FL to accelerate the model training by over-the-air aggregation, while relaxing the synchronization requirement of this aggregation technology.

\section{System Model}\label{sec:prelim}

\subsection{Federated Learning (FL)}\label{subsec:federated}

For ease of expression, some key notations in this paper are listed in Table \ref{tbl:natation}. We consider a $K$-class classification problem with a label space $\mathcal{C} = \{c_1,c_2,...,c_K\}$, and perform federated learning over a set of workers $\mathcal{V}=\{v_1,v_2,...,v_N\}$, with $|\mathcal{V}|=N>1$, in edge computing. 

Each worker $v_i$ trains a model on a local dataset with size $d_i$. The size of the data with label $c_k$ on worker $v_i$ is $d_i^k$, and $\sum_{c_k\in\mathcal{C}}d_i^k=d_i$.
The total data size on all workers is denoted as $D=\sum_{v_i\in \mathcal{V}}d_i$.
Let $\alpha_i=d_i/D$ and $\lambda_k=\sum_{v_i\in\mathcal{V}}d_i^k/D$ denote the proportion of worker $v_i$'s data size and class $c_k$'s data size to the total data size, respectively.
Let $(\mathbf{x}, y)$ and $\mathbf{w}$ denote a particular labeled sample and the model parameter, respectively.
For classification, we use the widely adopted cross-entropy loss function \cite{dunne1997pairing}:
\begin{equation}\label{eq:lossfuction}
F(\mathbf{w})\triangleq\sum_{c_k\in\mathcal{C}}-\lambda_k\mathbb{E}_{\mathbf{x}|y=c_k}[\log p_{k}(\mathbf{x},\mathbf{w})]\mbox{,}
\end{equation}
where $p_k(\mathbf{x},\mathbf{w})$ predicts the probability that the input $\mathbf{x}$ belongs to the class $c_k$ under parameter $\mathbf{w}$. Similarly, the loss function of worker $v_i$ is defined as
\begin{equation}\label{eq:filossfuction}
f_i(\mathbf{w})\triangleq\sum_{c_k\in\mathcal{C}}-\alpha_i^k\mathbb{E}_{\mathbf{x}|y=c_k}[\log p_{k}(\mathbf{x},\mathbf{w})]\mbox{,}
\end{equation}
where $\alpha_i^k=\frac{d_i^k}{d_i}$ denotes the proportion of the data size of class $c_k$ on worker $v_i$. Obviously, the global loss function satisfies
\begin{equation}\label{eq:loss}
F(\mathbf{w})=\sum_{v_i\in \mathcal{V}}\frac{d_i}{D}f_i(\mathbf{w})=\sum_{v_i\in \mathcal{V}}\alpha_if_i(\mathbf{w})\mbox{.}
\end{equation}
The learning problem is to obtain the optimal parameter vector $\mathbf{w}^*$ so as to minimize $F(\mathbf{w})$, \ie, $\mathbf{w}^*=\mathop{\argmin}_{\mathbf{w}} F(\mathbf{w})$.

\begin{table}[t]
    \setlength{\abovecaptionskip}{10pt}%
\setlength{\belowcaptionskip}{0pt}%
    \centering
    \captionof{table}{Key Notations.}\label{tbl:natation}
    \begin{tabular}{lll}
    \toprule
        Symbol & Semantics \\ \midrule\vspace{0.2em}
        $\mathcal{V}$ & \tabincell{l}{The set of workers $\{v_1, v_2, ..., v_N\}$}\\\vspace{0.2em}
        $\mathcal{V}_{j}$ & \makecell[l]{The set of workers in the $j$-th group } \\\vspace{0.2em}
        $\mathbf{V}$ & \tabincell{l}{The set of groups $\{\mathcal{V}_1, \mathcal{V}_2, ..., \mathcal{V}_M\}$} \\\vspace{0.2em}
        $d_i,D_j,D$  & \tabincell{l}{The data size of worker $v_i$/group $\mathcal{V}_j$/total} \\\vspace{0.2em}
        $d_i^k,D_j^k$  & The data size of class $c_k$ on worker $v_i$/group $\mathcal{V}_j$\\\vspace{0.2em}
        $\alpha_i,\beta_j,\lambda_k$ & \makecell[l]{The proportion of the data size of worker $v_i$ \\/group $\mathcal{V}_j$/class $c_k$ to the total data size} \\\vspace{0.2em}
        $\alpha_i^k,\beta_j^k$ & \makecell[l]{The proportion of the data size of class $c_k$ \\ on worker $v_i$/group $\mathcal{V}_j$} \\\vspace{0.2em}
        $F,f_i$ & The global/local loss function \\\vspace{0.2em}
        $\mathbf{\hat{w}}_t$ & The error-free global model at round $t$ \\\vspace{0.2em}
        $\mathbf{w}_t$ & The estimate of global model at round $t$ \\\vspace{0.2em}
        $\mathbf{w}_t^i$ & \makecell[l]{The local model on worker $v_i$ at round $t$} \\\vspace{0.2em}
        $\mathbf{y}_t$ & The received signal on PS at round $t$ \\\vspace{0.2em}
        $\mathbf{z}_t$ & The white Gaussian noise at round $t$ \\\vspace{0.2em}
        $\tau_t$ & The staleness at round $t$ \\\vspace{0.2em}
        $\sigma_t$ & The scaling factor at round $t$ \\\vspace{0.2em}
        $p_t^i$ & \makecell[l]{The transmit power of $v_i$ at round $t$} \\\vspace{0.2em}
        $E_t^i$ & The energy consumption of $v_i$ at round $t$ \\\vspace{0.2em}
        $\eta_t$ & The denoising factor at round $t$ \\
    \bottomrule
    \end{tabular}
    \vspace{-0.2cm}
\end{table}

\subsection{Grouping Asynchronous Federated Learning via Over-the-air Computation (Air-FedGA)}\label{subsec:REAFL}

We propose the AirComp-based Grouping Asynchronous FL mechanism, which is formally described in Alg. \ref{alg:serverprocess}.

\subsubsection{Worker Grouping}

Workers in $\mathcal{V}$ are organized into $M$ groups $\mathcal{V}_1,...,\mathcal{V}_M$, satisfying $\bigcup_{j=1}^M\mathcal{V}_j=\mathcal{V}$ and $\mathcal{V}_j\bigcap\mathcal{V}_{j'}=\varnothing,\forall j\neq j'$.
Let $D_j$ denote the sum of the data size of workers in group $\mathcal{V}_j$, \ie, $D_j=\sum_{v_i\in\mathcal{V}_j}d_i$. Then the proportion of the data size of group $\mathcal{V}_j$ to the total data size is $\beta_j=D_j/D$. Let $D_j^k$ denote the total size of data labeled as $c_k$ in group $\mathcal{V}_j$, \ie, $D_j^k=\sum_{v_i\in\mathcal{V}_j}d_i^k$. Then the proportion of the data size of class $c_k$ in group $\mathcal{V}_j$ is $\beta_j^k=D_j^k/D_j$.

\subsubsection{Local Training}\label{subsubsec:localtraining}

\begin{algorithm}[t]
\caption{Grouping Asynchronous Federated Learning via Over-the-Air Computation (Air-FedGA)}\label{alg:serverprocess}
\begin{algorithmic}[1]
\FOR {$j\in[1,M]$}
\STATE $r_j=0$
\ENDFOR
\FOR {$t=1$ to $T$}
\STATE \textbf{Processing at Each Worker $v_i$}\label{alg:line:ecstart}
\IF{receive $\mathbf{w}_{t-1}$ from the PS} \label{alg:line:reglobalstart}
\STATE Update local model $\mathbf{w}_{k}^i$ by Eq. \eqref{localupdate}
\STATE Send READY message to the PS \label{alg:line:reglobalend}
\ELSE
\STATE $\mathbf{w}_t^i=\mathbf{w}_{t-1}^i$ \label{alg:line:equallast}
\ENDIF
\IF {Receive EXECUTE message from the PS} \label{alg:line:reaggstart}
\STATE Transmit $\mathbf{w}_{t}^i$ simultaneously with all the other workers in group $\mathcal{V}_{j_t}$\label{alg:line:ecend}
\ENDIF \label{alg:line:reaggend}
\STATE \textbf{Processing at the Parameter Server}\label{alg:line:psstart}
\WHILE {\textbf{True}}
\IF {Receive READY message from worker $v_i$}\label{alg:line:rereadystart}
\STATE {$j=\arg\{j\in[1,M]|v_i\in\mathcal{V}_j\}$}
\STATE {$r_{j}=r_{j}+1$}
\IF {$r_{j}=|\mathcal{V}_{j}|$}
\STATE $j_t=j$\label{alg:line:sendaggstart}
\STATE {$r_{j_t}=0$}
\STATE {Send EXECUTE message to each $v_i\in\mathcal{V}_{j_t}$ }\label{alg:line:sendaggend}
\STATE {Receive signal $\mathbf{y}_t$ by over-the-air aggregation}
\STATE {Update global model $\mathbf{\hat{w}}_t$ according to Eq. \eqref{eq:estglobalmodel}} \label{alg:line:psend}
\STATE {Distribute $\mathbf{\hat{w}}_{t}$ to each worker $v_i\in \mathcal{V}_{j_t}$}
\STATE {\textbf{break}}
\ENDIF
\ENDIF\label{alg:line:rereadyend}
\ENDWHILE
\ENDFOR
\STATE {\textbf{return} global model $\mathbf{w}_T$ }
\end{algorithmic}
\end{algorithm}

Let $\mathcal{V}_{j_t}$ denote the group that participating in the global updating at round $t$.
If worker $v_i\notin\mathcal{V}_{j_t}$, it will not receive global model from the parameter server (PS) at round $t$, and its local model at round $t$ is equal to that at round $t-1$, \ie, $\mathbf{w}_t^i=\mathbf{w}_{t-1}^i$ (Line \ref{alg:line:equallast}). On the contrary, if worker $v_i\in\mathcal{V}_{j_t}$, it receives the global model $\mathbf{w}_{t-1}$ and performs local updating by
\begin{equation}\label{localupdate}
\mathbf{w}_t^i=\mathbf{w}_{t-1}-\gamma\nabla f_{i}(\mathbf{w}_{t-1})\mbox{,}
\end{equation}
where $\gamma$ is the learning rate (\ie, step size).
Let $\tau_t$ be the interval between the current round $t$ and the last received global model version by worker in group $\mathcal{V}_{j_t}$, called the \emph{staleness}. Thus, $\mathbf{w}_t^i$ is equal to $\mathbf{w}_{t-\tau_t}^i$, and $\mathbf{w}_{t-\tau_t}^i$ is trained from a previous version of the global model on $v_i$, \ie,
\begin{equation}\label{eq:wti}
\mathbf{w}_{t}^i=\mathbf{w}_{t-\tau_t}^i=\mathbf{w}_{t-\tau_t-1}-\gamma\nabla f_{i}(\mathbf{w}_{t-\tau_t-1})\mbox{.}
\end{equation}
Then, $v_i$ sends a READY message to the parameter server (Lines \ref{alg:line:reglobalstart}-\ref{alg:line:reglobalend}).

\subsubsection{Intra-group Alignment}\label{subsubsec:localtraining}

The parameter server maintains a set of variables $r_j,\forall j\in[1,M]$. Once the parameter server receives a READY message from a worker in group $\mathcal{V}_j$, $r_j$ increases by 1 (Lines \ref{alg:line:rereadystart}-\ref{alg:line:rereadyend}). If the parameter server has received all READY messages of worker in group $\mathcal{V}_j$, \ie, $r_j=|\mathcal{V}_j|$, it sends the EXECUTE messages to workers in $\mathcal{V}_j$ and reset $r_j$ as 0 (Lines \ref{alg:line:sendaggstart}-\ref{alg:line:sendaggend}).

\subsubsection{Grouping Asynchronous Aggregation}\label{subsubsec:groupasynchronous}

On receiving the EXECUTE message, worker $v_i$ transmits $\mathbf{w}_{t}^i$ and performs over-the-air aggregation simultaneously with all the other participating workers (Lines \ref{alg:line:reaggstart}-\ref{alg:line:reaggend}).
Let $h_t^i$ denote the wireless channel gain between worker $v_i$ and the parameter server at round $t$, which is assumed to remain unchanged within one communication round. Then the transmit power of $v_i$ at round $t$ is set as
\begin{equation}
p_t^i=\frac{d_i\sigma_t}{h_t^i}\mbox{,}
\end{equation}
where $\sigma_t$ is the power scaling factor at round $t$ determining the received SNR at the parameter server.
According to \cite{sun2021dynamic}, the transmission energy consumption on worker $v_i$ at round $t$ is given by
\begin{equation}\label{commenergy}
E_{t}^{i}=\|p_t^i\mathbf{w}_t^i\|_2^2\mbox{.}
\end{equation}
As a result, their local models are aggregated over-the-air for the parameter server.
If the aggregation is error-free, the global model can be obtained by
\begin{equation}\label{eq:globalupdate}
\mathbf{\hat{w}}_{t}=\left(1-\sum\nolimits_{v_i\in \mathcal{V}_{j_t}}\alpha_i\right)\mathbf{w}_{t-1}+\sum\nolimits_{v_i\in \mathcal{V}_{j_t}}\alpha_i\mathbf{w}_{t}^i\mbox{.}
\end{equation}
However, due to channel fading and noise, the received signal at the parameter server is given by
\begin{align}\label{eq:yt}
\mathbf{y}_t=\sum\nolimits_{v_i\in\mathcal{V}_{j_t}}p_t^ih_t^i\mathbf{w}_t^i+\mathbf{z}_t=\sum\nolimits_{v_i\in\mathcal{V}_{j_t}}d_i\sigma_t\mathbf{w}_t^i+\mathbf{z}_t\mbox{,}
\end{align}
where $\mathbf{z}_t$ is an additive white Gaussian noise (AWGN) vector with zero mean and variance $\sigma_0^2$. Therefore, the parameter server estimates the global model as
\begin{align}\label{eq:estglobalmodel}
\mathbf{w}_t=\left(1-\sum\nolimits_{v_i\in\mathcal{V}_{j_t}}\alpha_i\right)\mathbf{w}_{t-1}+\frac{\mathbf{y}_t}{D\sqrt{\eta_t}} \mbox{,}
\end{align}
where $\eta_t$ is the denoising factor at round $t$.
\begin{figure}[t]\centering
\includegraphics[width=0.49\textwidth]{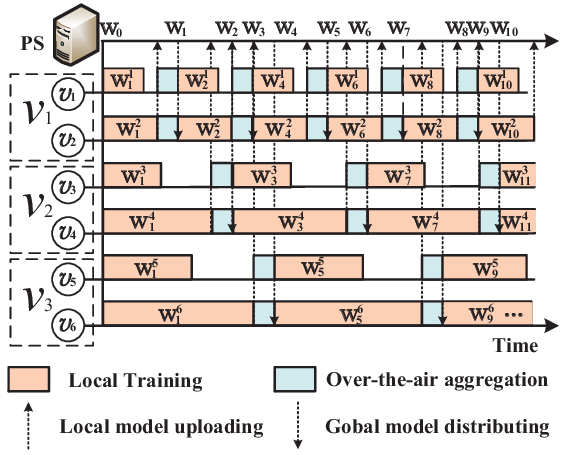}
\caption{The workflow of Air-FedGA.} \label{fig:intergroup}
\vspace{-2mm}
\end{figure}
After aggregation, the parameter server distributes the global model $\mathbf{w}_t$ to all workers in $\mathcal{V}_{j_t}$.

To visualize the procedure of Air-FedGA, we give an example in Fig. \ref{fig:intergroup}. There are 6 workers $v_1$-$v_6$ in the FL system, divided into 3 groups, \ie, $\mathcal{V}_1=\{v_1,v_2\}$, $\mathcal{V}_2=\{v_3,v_4\}$, $\mathcal{V}_3=\{v_5,v_6\}$ and $\mathbf{V}=\{\mathcal{V}_1,\mathcal{V}_2,\mathcal{V}_3\}$. For instance, workers $v_1$ and $v_2$ performs over-the-air aggregation simultaneously at round 1, \ie, $\mathbf{w}_1=(1-\alpha_1-\alpha_2)\mathbf{w}_0+\frac{\alpha_1\sigma_1\mathbf{w}_1^1+\alpha_2\sigma_1\mathbf{w}_1^2}{\sqrt{\eta_1}}+ \frac{\mathbf{z}_1}{D\sqrt{\eta_1}}$. Since workers $v_1$ and $v_2$ receive the global model $\mathbf{w}_0$ at round 1, the staleness $\tau_1=0$. For another instance, workers $v_5$ and $v_6$ performs over-the-air aggregation simultaneously at round 4, \ie, $\mathbf{w}_4=(1-\alpha_5-\alpha_6)\mathbf{w}_3+\frac{\alpha_5\sigma_4\mathbf{w}_4^5+\alpha_6\sigma_4\mathbf{w}_4^6}{\sqrt{\eta_4}}+ \frac{\mathbf{z}_4}{D\sqrt{\eta_4}}$. Since the last time workers $v_4$ and $v_6$ receive the global model $\mathbf{w}_0$ at round 1, $\mathbf{w}_4^5=\mathbf{w}_1^5$ and $\mathbf{w}_4^6=\mathbf{w}_1^6$, and the staleness $\tau_4=4-1=3$.

\section{Convergence Analysis}\label{sec:convergence}

\subsection{Assumptions}\label{subsec:assumptions}

We make the following commonly adopted assumptions \cite{cao2021optimized}\cite{cao2022transmission} on the loss functions $F_i,\forall v_i\in \mathcal{V}$.
\begin{assumption}[Smoothness]\label{ass:smoothness}
$F_i$ is L-smooth with $L > 0$, \ie, for $\forall \mathbf{w}_1, \mathbf{w}_2$, $F_i(\mathbf{w}_2)-F_i(\mathbf{w}_1)\le\langle \nabla F_i(\mathbf{w}_1),\mathbf{w}_2-\mathbf{w}_1\rangle+\frac{L}{2}{\|\mathbf{w}_2-\mathbf{w}_1\|}^2$.
\end{assumption}
\begin{assumption}[Strong convexity]\label{ass:strongconvexity}
$F_i$ is $\mu$-strongly convex with $\mu\ge 0$, \ie, for $\forall \mathbf{w}_1, \mathbf{w}_2$, $F_i(\mathbf{w}_2)-F_i(\mathbf{w}_1)\ge\langle \nabla F_i(\mathbf{w}_1), \mathbf{w}_2-\mathbf{w}_1 \rangle+\frac{\mu}{2}{\|\mathbf{w}_2-\mathbf{w}_1\|}^2$.
\end{assumption}
Note that models with convex and smooth loss functions (\eg, linear regression and SVM) satisfy Assumptions \ref{ass:smoothness} and \ref{ass:strongconvexity}. However, the evaluation results in Section \ref{sec:evaluation} demonstrate that our mechanism also performs well for models (\eg, CNN) with non-convex or non-smooth loss functions.
\begin{assumption}[Gradient bound]\label{ass:bounded}
The squared norm of gradients is uniformly bounded, \ie, $\forall k,\mathbf{w}$, $\|G_k(\mathbf{w})\|^2\le G^2$, where $G_k(\mathbf{w})=\nabla \mathbb{E}_{\mathbf{x}|y=c_k}[\log p_{k}(\mathbf{x},\mathbf{w})]$.
\end{assumption}
\begin{assumption}[Model bound]\label{ass:modelbound}
The squared norm of model size is uniformly bounded, \ie, $\|\mathbf{w}_t^i\|^2\le W_t^2$, $\forall v_i\in\mathcal{V},t\in[T]$.
\end{assumption}
Assumption \ref{ass:modelbound} is reasonable since the model size is relatively stable during federated learning process.

\subsection{Analysis of Convergence Bound}\label{subsec:convergencebound}

The earth mover distance (EMD) is applied to represent the difference of data distribution between two datasets \cite{zhao2018federated}.
We denote $\Lambda_j$ as the EMD between group $\mathcal{V}_j$'s dataset $\mathcal{D}_j$ and the global dataset $\mathcal{D}$, \ie,
\begin{align}
\Lambda_j=\mbox{EMD}(\mathcal{D},\mathcal{D}_j)=\sum\nolimits_{c_k\in\mathcal{C}}\|\lambda_k-\beta_j^k\|\mbox{.}
\end{align}
Before convergence analysis, we first state a key lemma for our statement.
For ease of expression, we denote $l_t=t-\tau_t-1\ge0 $ as the version of the received global model on $\mathcal{V}_{j_t}$ before the $t$th global aggregation. $\tau_{max}=\max_t\{\tau_t\}$ denotes the maximum staleness.
\begin{lemma}\label{lem:Q}
Let ${Q(t)}$ be a sequence of real numbers for $t\ge0$. $x$, $y$ and $z$ are three nonnegative constants, satisfying $x+y<1$. If $Q(t)\le xQ(t-1)+yQ(l_t)+z$, then
\begin{equation}\label{Vk}
Q(t)\le \rho^tQ(0)+\delta\mbox{,}
\end{equation}
where $\rho={(x+y)}^{\frac{1}{1+\tau_{max}}}$ and $\delta=\frac{z}{1-x-y}$.
\end{lemma}
Lemma \ref{lem:Q} can be proved by mathematical induction \cite{feyzmahdavian2014delayed}. Next, we derive the specific values of $x$, $y$ and $z$ in Air-FedGA and obtain the convergence bound by applying Lemma \ref{lem:Q}. For asynchronous aggregation among groups, we denote $\psi_j$ as the relative frequency of group $\mathcal{V}_j$ participating in the global aggregation, satisfying $\sum_{\mathcal{V}_j\in\mathbf{V}}\psi_j=1$.
\begin{theorem}\label{thm:convergence}
$\mathbf{w}_0$ is the initial global model. If $\frac{1}{2L}<\gamma<\frac{1}{L}$, after the over-the-air aggregation Eq. \eqref{eq:estglobalmodel} is performed $T$ times, the trained global model $\mathbf{w}_{T}$ satisfies
\begin{align}
&\mathbb{E}[F(\mathbf{w}_T)]-F(\mathbf{w}^*)\le\rho^T(F(\mathbf{w}_0)-F(\mathbf{w}^*))+\delta\mbox{,}
\end{align}
where $\rho=[1-(2\mu\gamma-\frac{\mu}{L})\sum_{\mathcal{V}_j\in \mathbf{V}}\psi_j\beta_j]^{\frac{1}{1+\tau_{max}}}\in(0,1)$, $\delta=\frac{\sum_{\mathcal{V}_j\in \mathbf{V}}\psi_j\beta_j(\gamma L\Lambda_j^2G^2+L^2\max_{t\in[1,T]}C_t)}
{(2\mu\gamma L-\mu)\sum_{\mathcal{V}_j\in \mathbf{V}}\psi_j\beta_j}$ and $C_t=(\frac{\sigma_t}{\sqrt{\eta_t}}-1)^2W_t^2+\frac{\sigma_0^2}{D_{j_t}^2\eta_t}$.
\end{theorem}
\begin{IEEEproof}
According to Eq. \eqref{eq:wti}, the local model
\begin{align}\notag
\mathbf{w}_{t}^i=&\mathbf{w}_{t-\tau_t}^i=\mathbf{w}_{t-\tau_t-1}-\gamma\nabla f_{i}(\mathbf{w}_{t-\tau_t-1})\\
=&\mathbf{w}_{l_t}-\gamma\nabla f_{i}(\mathbf{w}_{l_t})\mbox{.}
\end{align}
Let $\mathbf{w}_t^j=\sum_{v_i\in\mathcal{V}_{j_t}}\frac{d_i}{D_{j_t}}\mathbf{w}_t^i$ denote the group model of $\mathcal{V}_j$ at round $t$, we have
\begin{align}\notag
\mathbf{w}_{t}^j=&\sum_{v_i\in\mathcal{V}_{j_t}}\frac{d_i}{D_{j_t}}\mathbf{w}_{l_t}-\gamma\sum_{v_i\in\mathcal{V}_{j_t}}\frac{d_i}{D_{j_t}}\nabla f_{i}(\mathbf{w}_{l_t})\\
=&\mathbf{w}_{l_t}-\gamma\nabla F_{j}(\mathbf{w}_{l_t})\mbox{,}
\end{align}
where $F_{j}(\mathbf{w})=\sum_{v_i\in\mathcal{V}_j}\frac{d_i}{D_j}f_{i}(\mathbf{w})$ is the group loss function.
From Eqs. \eqref{eq:yt} and \eqref{eq:estglobalmodel},
\begin{align}\notag
\mathbf{w}_t=&\left(1-\sum\nolimits_{v_i\in\mathcal{V}_{j_t}}\frac{D_i}{D}\right)\mathbf{w}_{t-1}+\frac{\mathbf{y}_t}{D\sqrt{\eta_t}} \\\notag
=&\left(1-\sum\nolimits_{v_i\in\mathcal{V}_{j_t}}\alpha_i\right)\mathbf{w}_{t-1}+\frac{\sum_{v_i\in\mathcal{V}_{j_t}}d_i\sigma_t\mathbf{w}_t^i+\mathbf{z}_t}{D\sqrt{\eta_t}}\\\notag
=&(1-\beta_{j_t})\mathbf{w}_{t-1}+\frac{D_{j_t}}{D}\frac{\sum_{v_i\in\mathcal{V}_{j_t}}d_i\sigma_t\mathbf{w}_t^i+\mathbf{z}_t}{D_{j_t}\sqrt{\eta_t}}\\
=&(1-\beta_{j_t})\mathbf{w}_{t-1}+\beta_{j_t}\mathbf{\tilde{w}}_t^j\mbox{,}
\end{align}
where $\mathbf{\tilde{w}}_t^j=\sum_{v_i\in\mathcal{V}_{j_t}}\frac{d_i\sigma_t\mathbf{w}_t^i+\mathbf{z}_t}{D_{j_t}\sqrt{\eta_t}}$.
The group model aggregation error of $\mathcal{V}_{j_t}$ caused by the over-the-air aggregation at round $t$ is given by
\begin{align}\notag
\varepsilon_t^j=&\mathbf{\tilde{w}}_t^j-\mathbf{w}_t^j \\\notag
=&\sum_{v_i\in\mathcal{V}_{j_t}}\frac{d_i\sigma_t\mathbf{w}_t^i+\mathbf{z}_t}{D_{j_t}\sqrt{\eta_t}}-\sum_{v_i\in\mathcal{V}_{j_t}}\frac{d_i}{D_{j_t}}\mathbf{w}_t^i \\
=&\sum_{v_i\in\mathcal{V}_{j_t}}\frac{d_i}{D_{j_t}}\mathbf{w}_t^i\left(\frac{\sigma_t}{\sqrt{\eta_t}}-1\right)+\sum_{v_i\in\mathcal{V}_{j_t}}\frac{\mathbf{z}_t}{D_{j_t}\sqrt{\eta_t}}\mbox{.}
\end{align}
Since $F$ is convex and $\beta_{j_t}\in(0,1]$, we can deduce that
\begin{align}\notag
&F(\mathbf{w}_{t})-F(\mathbf{w}^*)\\\notag
=&F((1-\beta_{j_t})\mathbf{w}_{t-1}+\beta_{j_t}\mathbf{\tilde{w}}_t^j)-F(\mathbf{w}^*) \\\notag
\le& (1-\beta_{j_t})F(\mathbf{w}_{t-1})+\beta_{j_t}F(\mathbf{\tilde{w}}_t^j)-F(\mathbf{w}^*) \\\label{eq:Fdifference0}
=& (1-\beta_{j_t})(F(\mathbf{w}_{t-1})-F(\mathbf{w}^*))+\beta_{j_t}(F(\mathbf{\tilde{w}}_t^j)-F(\mathbf{w}^*))\mbox{.}
\end{align}
According to Assumption \ref{ass:smoothness}, it is obvious that $F$ is $L$-smooth, it follows
\begin{align}\notag
&F(\mathbf{\tilde{w}}_{t}^j)-F(\mathbf{w}^*)\\\notag
\le& F(\mathbf{w}_{l_t})-F(\mathbf{w}^*)+\langle\nabla F(\mathbf{w}_{l_t}),\mathbf{\tilde{w}}_{t}^j-\mathbf{w}_{l_t}\rangle+\frac{L}{2}\|\mathbf{\tilde{w}}_t^j-\mathbf{w}_{l_t}\|^2 \\\notag
=& F(\mathbf{w}_{l_t})-F(\mathbf{w}^*)+\langle\nabla F(\mathbf{w}_{l_t}),\mathbf{w}_{t}^j+\varepsilon_t^j-\mathbf{w}_{l_t}\rangle \\\notag
&+\frac{L}{2}\|\mathbf{w}_t^j+\varepsilon_t^j-\mathbf{w}_{l_t}\|^2 \\\notag
=& F(\mathbf{w}_{l_t})-F(\mathbf{w}^*)-\gamma\langle\nabla F(\mathbf{w}_{l_t}),\nabla F_j(\mathbf{w}_{l_t})\rangle\\\notag
&+\langle\nabla F(\mathbf{w}_{l_t}),\varepsilon_t^j\rangle+\frac{L\gamma^2}{2}\|\nabla F_j(\mathbf{w}_{l_t})\|^2+\frac{L\|\varepsilon_t^j\|^2}{2} \\\label{eq:Fdifference}
&-L\gamma\langle\varepsilon_t^j,\nabla F_{j}(\mathbf{w}_{l_t})\rangle\mbox{.}
\end{align}
By using the AM-GM Inequality, we have
\begin{align}\notag
&\langle\varepsilon_t^j,\nabla F(\mathbf{w}_{l_t})-L\gamma\nabla F_j(\mathbf{w}_{l_t})\rangle \\\notag
\le & \frac{L\|\varepsilon_t^j\|^2}{2}+\frac{\|\nabla F(\mathbf{w}_{l_t})-L\gamma\nabla f_i(\mathbf{w}_{l_t})\|^2}{2L} \\\notag
= & \frac{L\|\varepsilon_t^j\|^2}{2}+\frac{\|\nabla F(\mathbf{w}_{l_t})\|^2}{2L}+\frac{L\gamma^2\|\nabla F_j(\mathbf{w}_{l_t})\|^2}{2} \\\label{eq:epsilontemp}
&-\gamma\langle\nabla F(\mathbf{w}_{l_t}),\nabla F_j(\mathbf{w}_{l_t})\rangle\mbox{.}
\end{align}
Since $\gamma<\frac{1}{L}$, by taking Eq. \eqref{eq:epsilontemp} into Eq. \eqref{eq:Fdifference}, we deduce that
\begin{align}\notag
&F(\mathbf{\tilde{w}}_{t}^j)-F(\mathbf{w}^*)\\\notag
\le& F(\mathbf{w}_{l_t})-F(\mathbf{w}^*)-2\gamma\langle\nabla F(\mathbf{w}_{l_t}),\nabla F_j(\mathbf{w}_{l_t})\rangle\\\notag
&+\gamma^2L\|\nabla F_j(\mathbf{w}_{l_t})\|^2+L\|\varepsilon_t^j\|^2+\frac{\|\nabla F(\mathbf{w}_{l_t})\|^2}{2L}\\\notag
\le& F(\mathbf{w}_{l_t})-F(\mathbf{w}^*)+\gamma\|\nabla F(\mathbf{w}_{l_t})-\nabla F_j(\mathbf{w}_{l_t})\|^2 \\\label{eq:Fdifference2}
&-\gamma\|\nabla F(\mathbf{w}_{l_t})\|^2+L\|\varepsilon_t^j\|^2+\frac{\|\nabla F(\mathbf{w}_{l_t})\|^2}{2L}\mbox{.}
\end{align}
From Eq. \eqref{eq:lossfuction}, for $\forall\mathbf{w}$, its gradient over the global dataset is
\begin{align}\notag
\nabla F(\mathbf{w})=&\sum_{c_k\in\mathcal{C}}-\lambda_k\nabla\mathbb{E}_{\mathbf{x}|y=c_k}[\log p_{k}(\mathbf{x},\mathbf{w})]\\
=&\sum_{c_k\in\mathcal{C}}-\lambda_kG_k(\mathbf{w})\mbox{.}
\end{align}
Similarly, its gradient over dataset $\mathcal{D}_j$ is
\begin{align}
\nabla F_j(\mathbf{w})=\sum_{c_k\in\mathcal{C}}-\beta_j^kG_k(\mathbf{w})\mbox{.}
\end{align}
According to Assumption \ref{ass:bounded}, we have
\begin{align}\notag
\|\nabla F(\mathbf{w})-\nabla F_j(\mathbf{w})\|^2=& \|\sum_{c_k\in\mathcal{C}}(\lambda_k-\beta_j^k)G_k(\mathbf{w})\|^2 \\\label{eq:gradientdifference}
\le& \Lambda_j^2G^2\mbox{.}
\end{align}
Substituting $\mathbf{w}_{l_t}^j$ into Eq. \eqref{eq:gradientdifference}, we obtain that
\begin{align}\label{eq:censyngradientgap}
\|\nabla F(\mathbf{w}_{l_t}^j)-\nabla F_j(\mathbf{w}_{l_t}^j)\|^2\le \Lambda_j^2G^2\mbox{.}
\end{align}
According to Assumption \ref{ass:strongconvexity}, it is obvious that $F$ is $\mu$-strongly convex, it follows
\begin{align}
\|\nabla F(\mathbf{w}_{l_t})\|^2\ge 2\mu(F(\mathbf{w}_{l_t})-F(\mathbf{w}^*))\mbox{.}
\end{align}
Since $\gamma>\frac{1}{2L}$, we have
\begin{align}\label{eq:censynconvexF}
(-\gamma+\frac{1}{2L})\|\nabla F(\mathbf{w}_{l_t})\|^2\le (-2\mu\gamma+\frac{\mu}{L})(F(\mathbf{w}_{l_t})-F(\mathbf{w}^*))\mbox{.}
\end{align}
By taking Eqs. \eqref{eq:gradientdifference} and \eqref{eq:censynconvexF} into Eq. \eqref{eq:Fdifference2}, we deduce that
\begin{align}\notag
&F(\mathbf{\tilde{w}}_t^j)-F(\mathbf{w}^*)\\\label{eq:Fdifference3}
\le& (1-2\mu\gamma+\frac{\mu}{L})(F(\mathbf{w}_{l_t})-F(\mathbf{w}^*))+\gamma\Lambda_j^2G^2+L\|\varepsilon_t^j\|^2\mbox{.}
\end{align}
By taking Eq. \eqref{eq:Fdifference3} into Eq. \eqref{eq:Fdifference0} and taking the expectation, we have
\begin{align}\notag
&\mathbb{E}[F(\mathbf{w}_{t})]-F(\mathbf{w}^*)\\\notag
\le& (1-\sum_{\mathcal{V}_j\in \mathbf{V}}\psi_j\beta_j)(F(\mathbf{w}_{t-1})-F(\mathbf{w}^*))\\\notag
&+(1-2\mu\gamma+\frac{\mu}{L})\sum_{\mathcal{V}_j\in \mathbf{V}}\psi_j\beta_j(F(\mathbf{w}_{l_t})-F(\mathbf{w}^*)) \\\label{con:Fw-Fw}
&+\sum_{\mathcal{V}_j\in \mathbf{V}}\psi_j\beta_j\left(\gamma\Lambda_j^2G^2+L\max_{t\in[1,T]}C_t\right)\mbox{.}
\end{align}
where
\begin{align}\label{eq:Ct}
C_t=\left(\frac{\sigma_t}{\sqrt{\eta_t}}-1\right)^2W_t^2+\frac{\sigma_0^2}{D_{j_t}^2\eta_t}
\end{align}
Let $Q(t)\triangleq\mathbb{E}[F(\mathbf{w}_{t})]-F(\mathbf{w}^*)$. Then $F(\mathbf{w}_{t-1})-F(\mathbf{w}^*)=Q(t-1)$ and $F(\mathbf{w}_{l_t})-F(\mathbf{w}^*)=Q(l_t)$. The recursive relation is transformed into
\begin{align}\notag
Q(t)\le& \underbrace{(1-\sum_{\mathcal{V}_j\in \mathbf{V}}\psi_j\beta_j)}_{x_t}Q(t-1)\\\notag
&+\underbrace{(1-2\mu\gamma+\frac{\mu}{L})\sum_{\mathcal{V}_j\in \mathbf{V}}\psi_j\beta_j}_{y_t}Q(l_t) \\
&+\underbrace{\sum_{\mathcal{V}_j\in \mathbf{V}}\psi_j\beta_j(\gamma\Lambda_j^2G^2+L\max_{t\in[1,T]}C_t)}_{z_t}\mbox{.}
\end{align}
According to Lemma \ref{lem:Q}, if $\mu\gamma\sum_{\mathcal{V}_j\in\mathbf{V}}\psi_j\beta_j\in(0,1)$, then
\begin{align}
&\mathbb{E}[F(\mathbf{w}_T)]-F(\mathbf{w}^*)\le\rho^T(F(\mathbf{w}_0)-F(\mathbf{w}^*))+\delta\mbox{,}
\end{align}
where $\rho=[1-(2\mu\gamma-\frac{\mu}{L})\sum_{\mathcal{V}_j\in \mathbf{V}}\psi_j\beta_j]^{\frac{1}{1+\tau_{max}}}\in(0,1)$ and $\delta=\frac{\sum_{\mathcal{V}_j\in \mathbf{V}}\psi_j\beta_j(\gamma L\Lambda_j^2G^2+L^2\max_{t\in[1,T]}C_t)}
{(2\mu\gamma L-\mu)\sum_{\mathcal{V}_j\in \mathbf{V}}\psi_j\beta_j}$.
\end{IEEEproof}

\subsection{Discussions}\label{subsec:discussions}

We can draw some meaningful corollaries from Theorem \ref{thm:convergence}.
\begin{corollary}\label{cor:distribution}
The greater the degree of data Non-IID among groups, the larger the value of $\Lambda_j$ for each group $\mathcal{V}_j$, and the higher residual error $\delta$. Given IID data among groups, then $\Lambda_j=0$ for $\forall \mathcal{V}_j\in\mathbf{V}$, the residual $\delta$ can be reduced.
\end{corollary}
Corollary \ref{cor:distribution} shows that workers can be organized to make the data distribution among groups close to IID, so as to reduce the residual error $\delta$ and improve training performance.
\begin{corollary}\label{cor:taumax}
The convergence factor $\rho$ decreases as the upper bound of staleness $\tau_{max}$ decreases. $\tau_{max}$ depends partly on the number $M$ of groups. For example, if $M=1$, then $\tau_{max}=0$, and $\rho$ takes the minimum value.
\end{corollary}

Corollary \ref{cor:taumax} shows that we can decrease the convergence factor $\rho$ by decreasing the number $M$ of groups. However, it does not mean a short convergence time, because the completion time of a single round depends on the worker with the maximum completion time within a group. Consequently, less groups will result in a longer completion time of a single round. Accordingly, it is a significant problem to determine the group strategy to achieve better training performance, which will be elaborated on in Section \ref{sec:algorithm}.

\section{Problem Formulation and Algorithm Description}\label{sec:algorithm}

\subsection{Problem Formulation}\label{subsec:problem}

To exploit AirComp for low-latency model aggregation \cite{zhu2020broadband}, the model aggregation time is calculated as
\begin{equation}
L^u=\frac{q}{R}L^s\mbox{,}
\end{equation}
where $q$ is the dimension of the trained model, $R$ is the number of sub-channels, and $L^s$ is the symbol duration of an OFDM symbol.
Let $x_{i,j}$ denote the indicator for whether worker $v_i$ belongs to group $\mathcal{V}_j$ or not. The grouping strategy in the whole system is denoted as $\boldsymbol{x}=\{x_{i,j}\}_{v_i\in\mathcal{V},\mathcal{V}_j\in\mathbf{V}}$. Let $l_i$ denote the local training time on worker $v_i$, which is assumed to be estimated by the historical measurements. $\Delta l=\max_{v_i\in\mathcal{V}}\{l_i\}-\min_{v_i\in\mathcal{V}}\{l_i\}$ is the difference between the maximum and minimum local training time of workers in $\mathcal{V}$.
The time for all workers in group $\mathcal{V}_j$ to complete local training is determined by the worker with the longest training time. Then the completion time for $\mathcal{V}_j$ to complete local training and model uploading via over-the-air aggregation is calculated as
\begin{align}
L_j=\max_{v_i\in\mathcal{V}_j}\{l_i\}+L^u\mbox{.}
\end{align}
Therefore, the number of updates that group $\mathcal{V}_j$ participates in per unit time is $\frac{1}{L_j}$.
Since all groups participate in global updating asynchronously, the average completion time of one round is estimated as
\begin{align}
\overline{L}\approx\frac{1}{\frac{1}{L_1}+\frac{1}{L_2}+...+\frac{1}{L_M}}\mbox{.}
\end{align}
Thus, we formulate our problem as follows:
\begin{subequations}
\begin{align}
\textbf{(P1)}:&\min_{\boldsymbol{\sigma},\boldsymbol\eta,\boldsymbol{x}} \overline{L}T \\
{\st} \quad &F(\mathbf{w}_T)\le F(\mathbf{w}^*)+\varepsilon \label{eq:12}\\
&E_t^i\le \hat{E}^i, & \forall v_i\in\mathcal{V},t\in[T] \label{eq:13}\\
&L_j-L^u-l_i\le \xi\Delta l, & \forall v_i\in\mathcal{V}_j,\mathcal{V}_j\in\mathbf{V} \label{eq:14}\\
&x_{i,j}\in\{0,1\}, &v_i\in\mathcal{V},\mathcal{V}_j\in\mathbf{V} \label{eq:15}
\end{align}
\end{subequations}
The first inequality \eqref{eq:12} represents that the global model will converge after $T$ rounds, where $\varepsilon$ is the convergence threshold to guarantee the training accuracy. The second set of inequalities \eqref{eq:13} represent that each worker $v_i$ is subject to a maximum energy budget $\hat{E}^i$ at each round $t$. The third set of inequalities \eqref{eq:14} ensures the local training time of workers within each group is similar (\eg, $\xi=0.3$).
Our target is to determine the power scaling factors $\boldsymbol{\sigma}=\{\sigma_t|t\in[T]\}$, the denoising factors $\boldsymbol{\eta}=\{\eta_t|t\in[T]\}$ and the grouping strategy $\boldsymbol{x}$ to minimize the total training time, \ie,  $\min_{\boldsymbol{\sigma},\boldsymbol{\eta},\boldsymbol{x}} \overline{L}T$.

Theorem \ref{thm:convergence} provides the convergence bound of the global model after $T$ rounds. To satisfy the constraint in Eq. \eqref{eq:12}, we take the upper bound of $\mathbb{E}[F(\mathbf{w}_T)]-F(\mathbf{w}^*)$
less than $\varepsilon$,
\begin{align}
\rho^T(F(\mathbf{w}_0)-F(\mathbf{w}^*))+\delta\le\varepsilon\mbox{,}
\end{align}
where $\rho=[1-(2\mu\gamma-\frac{\mu}{L})\sum_{\mathcal{V}_j\in \mathbf{V}}\psi_j\beta_j]^{\frac{1}{1+\tau_{max}}}$.
We define that $A\triangleq\frac{\varepsilon-\delta}{F(\mathbf{w}_0)-F(\mathbf{w}^*)}$ and $B\triangleq1-(2\mu\gamma-\frac{\mu}{L})\sum_{\mathcal{V}_j\in \mathbf{V}}\psi_j\beta_j\in(0,1)$, then it holds that
\begin{align}
T \ge (1+\tau_{max})\log_{B}A\mbox{,}
\end{align}
It is obvious that the group with the largest staleness factor $\tau_{max}$ is also the group with the longest completion time. Therefore, $\tau_{max}$ can be estimated as
\begin{align}
\hat{\tau}_{max}=L_j^{max}\sum_{\mathcal{V}_j\in\mathbf{V}}\frac{1}{L_j}
\end{align}
Thus we convert the original problem $\textbf{P1}$ to the following optimization problem:
\begin{subequations}
\begin{align}\label{eq21}
\textbf{(P2)}:&\min_{\boldsymbol{\sigma},\boldsymbol{\eta},\boldsymbol{x}} \overline{L}(1+\hat{\tau}_{max})\log_{B}A \\
{\st} \quad &\eqref{eq:13},\eqref{eq:14} \mbox{ and } \eqref{eq:15}\mbox{.}
\end{align}
\end{subequations}

\subsection{Power Control}\label{subsec:powercontrol}

Since the power scaling factor $\sigma_t$ and the denoising factor $\eta_t$ at round $t$ are only related to $C_t=\left(\frac{\sigma_t}{\sqrt{\eta_t}}-1\right)^2W_t^2+\frac{\sigma_0^2}{D_{j_t}^2\eta_t}$ as indicated in Eq. \eqref{eq:Ct}, and are independent of the remaining terms in the optimization objective in \textbf{P2}, we decouple the process of solving for $\sigma_t$ and $\eta_t$ from \textbf{P2}. Specifically, we determine $\sigma_t$ and $\eta_t$ to minimize $C_t$, \ie,
\begin{subequations}
\begin{align}\label{eq21}
\textbf{(P3)}:&\min_{\sigma_t,\eta_t} C_t \\
{\st} \quad &E_t^i\le \hat{E}^i, & \forall v_i\in\mathcal{V},t\in[T] \label{eq:31}
\end{align}
\end{subequations}
Note that $\sigma_t$ and $\eta_t$ are coupled in \textbf{P3}. Therefore, we address \textbf{P3} by adopting the alternating optimization method \cite{cao2022transmission}, which is formally described in Alg. \ref{alg:power}. The main idea is to alternately fix $\sigma_t$/$\eta_t$ and determine the value of $\eta_t$/$\sigma_t$ to optimize $C_t$. After several iterations, convergent $\sigma_t^*$ and $\eta_t^*$ are obtained.

\begin{algorithm}[t]
\renewcommand{\algorithmicrequire}{\textbf{Input:}}
\renewcommand{\algorithmicensure}{\textbf{Output:}}
\caption{Iterative Algorithm for Power Control}\label{alg:power}
\begin{algorithmic}[1]
\REQUIRE Initial scaling factor $\sigma_t$
\ENSURE convergent $\sigma_t^*$ and $\eta_t^*$
\WHILE {$\frac{|\sigma_t^*-\sigma_t|}{\sigma_t^*}>\theta$ or $\frac{|\eta_t^*-\eta_t|}{\eta_t^*}>\theta$}
\STATE {$\sigma_t^*=\sigma_t$, $\eta_t^*=\eta_t$}
\STATE {$\eta_t=\left(\frac{\sigma_t^2W_t^2+\frac{\sigma_0^2}{D_{j_t}^2}}{\sigma_tW_t^2}\right)^2$}
\STATE {$\sigma_t=\min\{\sqrt{\eta_t}\}\bigcup\left\{\left.\frac{h_t^i\sqrt{\hat{E}^i}}{d_iW_t}\right|\forall v_i\in\mathcal{V}\right\}$}
\ENDWHILE
\STATE {$\sigma_t^*=\sigma_t$, $\eta_t^*=\eta_t$}
\STATE \textbf{return} convergent $\sigma_t^*$ and $\eta_t^*$
\end{algorithmic}
\end{algorithm}

At each iteration, we first optimize the denoising factors $\eta_t$ under given scaling factors $\sigma_t$. Let $\hat{\eta}_t=\frac{1}{\sqrt{\eta_t}}$, Eq. \eqref{eq:Ct} is transformed to
\begin{align}
C_t=(\sigma_t\hat{\eta}_t-1)^2W_t^2+\frac{\sigma_0^2\hat{\eta}_t^2}{D_{j_t}^2}\mbox{.}
\end{align}
The partial derivative of $C_t$ with respect to $\hat{\eta}_t$ is calculated as
\begin{align}
\frac{\partial C_t}{\partial \hat{\eta}_t}=2\left(\sigma_t^2\hat{\eta}_tW_t^2-\sigma_tW_t^2+\frac{\sigma_0^2}{D_{j_t}^2}\hat{\eta}_t\right)\mbox{.}
\end{align}
Since $C_t$ is convex with respect to $\hat{\eta}_t$, the necessary condition for minimization is given by setting the partial derivative to zero, \ie, $\frac{\partial C_t}{\partial \hat{\eta}_t}=0$. Solving this equation yields the optimal value of $\hat{\eta}_t$ as $\hat{\eta}_t=\frac{\sigma_tW_t^2}{\sigma_t^2W_t^2+\frac{\sigma_0^2}{D_{j_t}^2}}$, \ie,
\begin{align}
\eta_t=\left(\frac{\sigma_t^2W_t^2+\frac{\sigma_0^2}{D_{j_t}^2}}{\sigma_tW_t^2}\right)^2\mbox{.}
\end{align}

Next, we optimize the scaling factor $\sigma_t$ under given denoising factor $\eta_t$.
On the one hand, the partial derivative of $C_t$ with respect to $\sigma_t$ is calculated as
\begin{align}
\frac{\partial C_t}{\partial \sigma_t}=&2W_t^2\left(\frac{\sigma_t}{\eta_t}-\frac{1}{\sqrt{\eta_t}}\right) \mbox{.}
\end{align}
Given that $C_t$ is convex with respect to $\sigma_t$, we similarly set $\frac{\partial C_t}{\partial \sigma_t}=0$. Solving it shows that $C_t$ is minimized when $\sigma_t=\sqrt{\eta_t}$, provided that this value lies within the feasible region.
On the other hand, from Assumption \ref{ass:modelbound} and constraints Eq. \eqref{eq:31}, for $\forall v_i\in\mathcal{V}$, we have the following inequality:
\begin{align}
E_t^i=\|p_t^i\mathbf{w}_t^i\|_2^2\le\left(\frac{d_i\sigma_t}{h_t^i}\right)^2W_t^2\le \hat{E}^i\mbox{.}
\end{align}
This leads to the bound $\sigma_t\le\frac{h_t^i\sqrt{\hat{E}^i}}{d_iW_t}$ for $\forall v_i\in\mathcal{V}$. Therefore, if $\eta_t$ is given, $C_t$ can be minimized when
\begin{align}
\sigma_t=\min\{\sqrt{\eta_t}\}\bigcup\left\{\left.\frac{h_t^i\sqrt{\hat{E}^i}}{d_iW_t}\right|\forall v_i\in\mathcal{V}\right\}\mbox{.}
\end{align}

At last, with a given threshold $\theta$, the convergent $\sigma_t^*$ and $\eta_t^*$ can be obtained  by alternate optimization for iterations.

\subsection{Worker Grouping Algorithm}\label{subsec:groupingalgorithm}

\begin{algorithm}[t]
\renewcommand{\algorithmicrequire}{\textbf{Input:}}
\renewcommand{\algorithmicensure}{\textbf{Output:}}
\caption{Worker Grouping Algorithm for Air-FedGA}\label{alg:grouping}
\begin{algorithmic}[1]
\REQUIRE Data size $d_i$, $d_i^k$, $\forall v_i\in\mathcal{V}$, $\forall c_k\in\mathcal{C}$
\ENSURE Final grouping strategy $\boldsymbol{x}$
\STATE {Group set $\mathbf{V}=\varnothing$}
\STATE $M=1$
\STATE Sort each worker $v_i\in\mathcal{V}$ in descending order by $D_i$ as $\mathcal{Q}$\label{line:ccinitialend}
\FOR{each $v_i\in\mathcal{Q}$}
\STATE $\mathcal{L}_{temp}=+\infty$
\FOR{each $\mathcal{V}_j\in\mathbf{V}\cup\mathcal{V}_M$} \label{line:traversestart}
\STATE $x_{i,j}=1$ \label{line:xij1}
\IF {$\mathcal{L}(\boldsymbol{x}) < \mathcal{L}_{temp}$ \textbf{and} $L_j(\boldsymbol{x})-L^u-l_i\le \xi\Delta l$ } \label{line:Ljudgestart}
\STATE $\mathcal{L}_{temp}=\mathcal{L}(\boldsymbol{x})$
\STATE $j^*=j$
\ENDIF\label{line:Ljudgeend}
\STATE $x_{i,j}=0$
\ENDFOR
\IF {$j^*==M$}\label{line:creatVMstart}
\STATE $\mathbf{V}=\mathbf{V}\cup\mathcal{V}_M$
\STATE $M=M+1$
\ENDIF\label{line:creatVMend}
\STATE $x_{i,j^*}=1$ \label{line:traverseend}
\ENDFOR
\STATE \textbf{return} final grouping strategy $\boldsymbol{x}$
\end{algorithmic}
\end{algorithm}

After the scaling factors $\boldsymbol{\sigma}^*$ and the denoising factors $\boldsymbol{\eta}^*$ are determined, the parameters $\delta^*$ and $A^*$ are accordingly determined. The problem $\textbf{P2}$ can be converted to the following optimization problem:
\begin{subequations}
\begin{align}\label{eq21}
\textbf{(P4)}:&\min_{\boldsymbol{x}} \overline{L}(1+\hat{\tau}_{max})\log_{B}A \\
{\st}\quad & \eqref{eq:14} \mbox{ and } \eqref{eq:15}\mbox{.}
\end{align}
\end{subequations}

Without causing ambiguity, we denote $\mathcal{L}(\boldsymbol{x})=\overline{L}(1+\hat{\tau}_{max})\log_{B}A^*$ and $L_j(\boldsymbol{x})$ as the value of objective and $L_j$ under strategy $\boldsymbol{x}$, respectively.
We introduce the worker grouping algorithm to solve \textbf{P4}, which is formally described in Alg. \ref{alg:grouping}. There are two difficulties in solving \textbf{P4}: first, it is difficult to confirm the number $M$ of groups. Second, even after the number of groups is determined, the search space of the group strategy is unacceptable $O(M^N)$ if an exhaustive method is adopted. Therefore, we adopt a greedy-based method to solve \textbf{P4}.

The main idea of our algorithm is to determine which group each worker belongs to one by one, so as to minimize the current objective function $\mathcal{L}(\boldsymbol{x})$. Specifically, we first sort the workers in the descending order of their data sizes as a queue $\mathcal{Q}$ (Line \ref{line:ccinitialend}). Note that the purpose of sorting the workers is to make the algorithm preferentially traverse the workers with more data. In this way, the performance of the algorithm is usually better than traversing in random order. Then, for each worker $v_i\in\mathcal{Q}$, we attempt to organize it to each group $\mathcal{V}_j\in\mathcal{V}$ or individually as a new group $\mathcal{V}_M$, \ie, set $x_{i,j}=1$ (Line \ref{line:xij1}), and calculate the values of current $\mathcal{L}(\boldsymbol{x})$. We traverse all group $\mathcal{V}_j\in\mathbf{V}\cup\mathcal{V}_M$ and organize worker $v_i$ to the aggregator $\mathcal{V}_{j^*}$ that minimizes the value of $\mathcal{L}(\boldsymbol{x})$ (Lines \ref{line:traversestart}-\ref{line:traverseend}). In particular, if a single $v_i$ alone as a group minimizes $\mathcal{L}(\boldsymbol{x})$, a separate group $\mathcal{V}_M$ is created for it (Lines \ref{line:creatVMstart}-\ref{line:creatVMend}). Meanwhile, the constrains \eqref{eq:14}, \ie, $L_j(\boldsymbol{x})-L^u-l_i\le \xi\Delta l$, $\forall v_i\in\mathcal{V}_j$ should be guaranteed in this process (Line \ref{line:Ljudgestart}).
In the worst case, with each worker as a group, the time complexity is $O(N^2)$. However, given that $N$ is relatively small, the running time of worker grouping algorithm is negligible compared to the model training time.


\section{Performance Evaluation}\label{sec:evaluation}

\subsection{System Setup}\label{subsec:setup}

We use PyTorch to simulate a large-scale federated learning system, which consists of one parameter server and 100 workers. Each worker simulates an individual machine and trains a local model on its own dataset. We conduct our experiments on a deep learning workstation with a 10-core Intel Xeon CPU (Silver 4210R) and 4 NVIDIA GeForce RTX 3090 GPUs with 24GB GDDR6X. The system environment is Ubuntu 22.04, CUDA v11.7, and cuDNN v8.5.0.

\subsubsection{Models and Datasets}\label{subsubsec:modelsanddatasets}

We adopt three real-world classical datasets to conduct extensive experiments:

\begin{itemize}
    \item \textbf{MNIST \cite{lecun1998gradient}} comprises a collection of handwritten digits (from `0' to `9'), includes 60,000 training samples and 10,000 testing samples.
    \item \textbf{CIFAR-10 \cite{krizhevsky2009learning}} is composed of 60,000 color images, which is divided into 10 classes, each containing 6,000 images. The dataset is further split into 50,000 training images and 10,000 test images.
    \item \textbf{ImageNet-100} is a subset of dataset ImageNet \cite{russakovsky2015imagenet}, which contains 1,281,167 training images, 50,000 validation images and 100,000 test images, spread across 1,000 categories. To accommodate the limited resources of edge clients, we construct a subset of ImageNet by randomly selecting the samples of 100 out of 1000 categories.
\end{itemize}

To implement the Non-IID data among workers, we adopt the label skewed method to partition dataset \cite{ma2024feduc}. Specifically, the data in MNIST labeled as `0' are distributed to workers $v_1$-$v_{10}$, labeled as `1' are distributed to workers $v_{11}$-$v_{20}$,..., and labeled as `9' are distributed to workers $v_{91}$-$v_{100}$.

Three different models with distinct structures are implemented on the aforementioned datasets:

\begin{itemize}
    \item \textbf{LR \cite{hosmer2013applied} on MNIST.} The logistic regression (LR in short), which is constructed of a fully connected network with two hidden layers with 512 units, is adopted for the MNIST dataset.
    \item \textbf{CNN \cite{shalev2014understanding} on MNIST and CIFAR-10.} The plain CNN models are tailored for the MNIST and CIFAR-10. 1) For MNIST: It consists of two 5 $\times$ 5 convolution layers (20, 50 channels), two fully-connected layers with 800 and 500 units, and a softmax layer with 10 units. 2) For CIFAR-10: The CNN consists of two 5 $\times$ 5 convolution layers (32, 64 channels), two fully-connected layers with 1600 and 512 units, and a softmax layer with 10 units.
    \item \textbf{VGG-16 \cite{simonyan2014very} on ImageNet-100.} The VGG-16 model, which consists of 13 convolutional layers with a kernel of 3 $\times$ 3, followed by two dense layers and a softmax output layer, is adopted for the ImageNet-100 dataset.
\end{itemize}

\subsubsection{Simulation of Edge Heterogeneity}\label{subsubsec:testbed}

Let $\hat{l}_i$ denote the actual local training time on $v_i$. Due to the limitations of training resources and the large-scale scenario, we actually deploy the virtual workers $v_1$-$v_{100}$ on the single workstation for experimentation, their local training times are roughly equal, \ie, $\hat{l}_1\approx \hat{l}_2\approx...\approx\hat{l}_{100}$. To simulate the edge heterogeneity, we introduce a scaling factor $\kappa_i$ for each worker $v_i$, which is a random float number drawn uniformly from [1,10]. Then, we set the local training time on worker $v_i$ as $l_i=\kappa_i\hat{l}_i$, which means that after completing local training, $v_i$ waits for 0-9 times before sending the READY message to the parameter server. This adjusted time $l_i$ is then used to calculate its training completion time and recorded in a dynamically maintained list, $\mathbf{L}$. By monitoring the training completion times of all workers recorded in $\mathbf{L}$, we determine when each group performs over-the-air aggregations.
Additionally, we set the bandwidth $B=1\mbox{MHz}$, the noisy variance $\sigma_0^2=1\mbox{W}$, and the energy constraints $\hat{E}^i=10\mbox{J}$ for each worker in each round.

\subsubsection{Benchmarks and Metrics}\label{subsubsec:metric}

To highlight the benefits of applying AirComp to asynchronous federated learning, we evaluate our \textbf{Air-FedGA} mechanism against two OMA-based mechanisms and two AirComp-based mechanisms.

\begin{itemize}
    \item \textbf{FedAvg\cite{mcmahan2017communication-efficient}:} A classic OMA-based synchronous mechanism, where all workers participating in each round of global aggregation.
    \item \textbf{TiFL\cite{chai2020tifl}:} An OMA-based group asynchronous mechanism, which organize workers into groups according to their communication time with the parameter server, and perform global updating asynchronously among groups.
    \item \textbf{Air-FedAvg\cite{cao2022transmission}:} The version of using AirComp technique to implement the FedAvg mechanism with optimal power control.
    \item \textbf{Dynamic\cite{sun2021dynamic}:} An AirComp-based synchronous solution, which dynamically selects a subset of workers for each round of global aggregation, while the rest remain idle.
\end{itemize}

To evaluate the training performance, we adopt the following performance metrics. 1) \emph{Loss Function} reflects the training process of the model and whether convergence has been achieved. 2) \emph{Accuracy} is the most common performance metric in classification problems, which is defined as the proportion of right data classified by the model to all test data. 3) \emph{Training Time} is adopted to measure the training rate.

\subsection{Evaluation Results}\label{subsec:result}

In this section, we first compare the performance of our proposed Air-FedGA with other benchmarks in terms of loss function and accuracy. We then demonstrate the advantages of Air-FedGA in handling heterogeneity, handling Non-IID data, and scalability.

\subsubsection{Loss Function and Accuracy}\label{subsubsec:convergencetime}

\begin{figure}[t]
\includegraphics[width=0.49\linewidth,height=3.6cm]{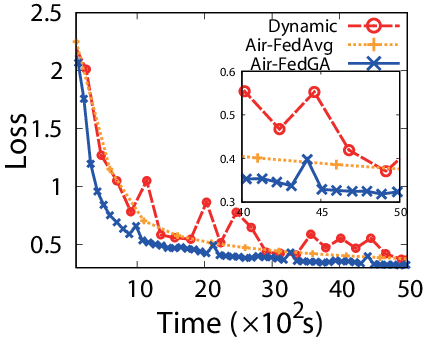}
\includegraphics[width=0.49\linewidth,height=3.6cm]{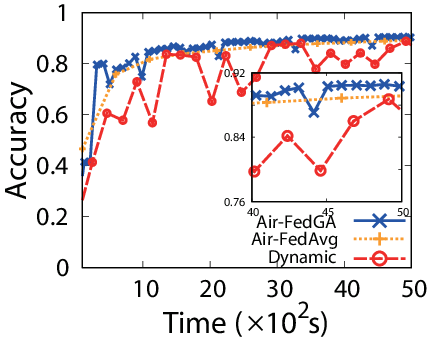}
\small{\caption{Loss/Accuracy vs. Time (LR on MNIST). \textit{Left}: Loss; \textit{Right}: Accuracy.}\label{fig_LR_mnist}}
\vspace{-2mm}
\end{figure}
\begin{figure}[t]
\includegraphics[width=0.49\linewidth,height=3.6cm]{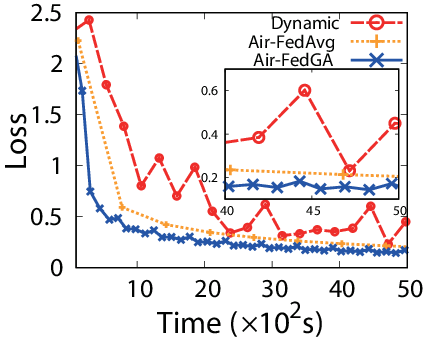}
\includegraphics[width=0.49\linewidth,height=3.6cm]{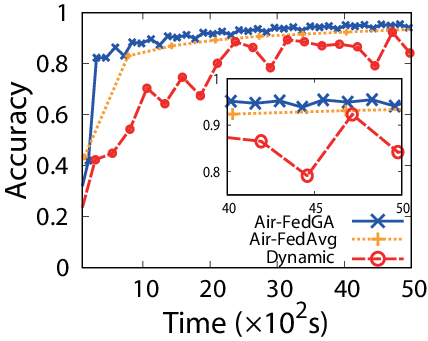}
\small{\caption{Loss/Accuracy vs. Time (CNN on MNIST). \textit{Left}: Loss; \textit{Right}: Accuracy.}\label{fig_CNN_mnist}}
\vspace{-2mm}
\end{figure}
\begin{figure}[t]
\includegraphics[width=0.49\linewidth,height=3.6cm]{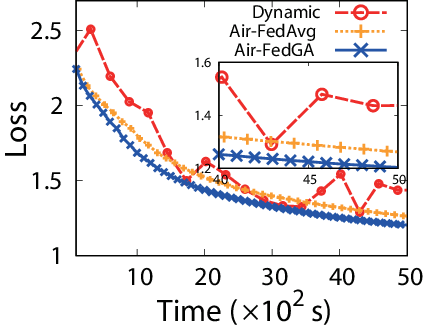}
\includegraphics[width=0.49\linewidth,height=3.6cm]{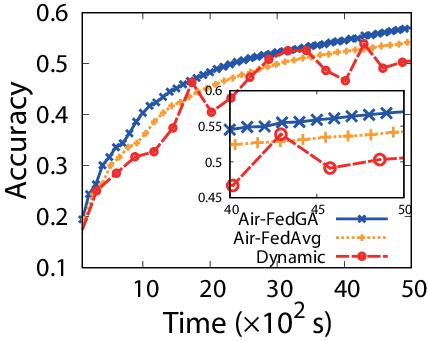}
\small{\caption{Loss/Accuracy vs. Time (CNN on CIFAR-10). \textit{Left}: Loss; \textit{Right}: Accuracy.}\label{fig_CNN_cifar}}
\vspace{-2mm}
\end{figure}
\begin{figure}[t]
\includegraphics[width=0.49\linewidth,height=3.6cm]{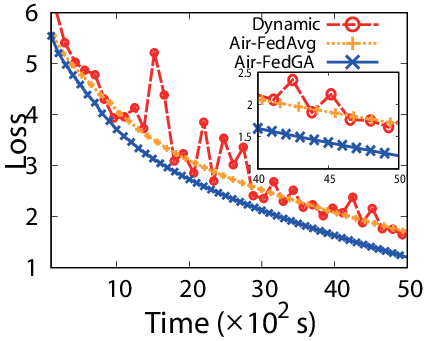}
\includegraphics[width=0.49\linewidth,height=3.6cm]{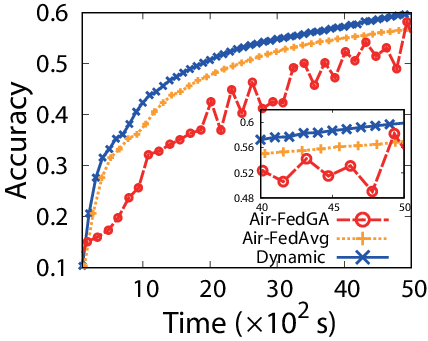}
\small{\caption{Loss/Accuracy vs. Time (VGG-16 on ImageNet-100). \textit{Left}: Loss; \textit{Right}: Accuracy.}\label{fig_VGG_ImageNet}}
\vspace{-2mm}
\end{figure}

Figs. \ref{fig_LR_mnist}-\ref{fig_VGG_ImageNet} illustrate the loss and accuracy curves over time for three models trained on three datasets. As a control experiment, our Air-FedGA is compared with two AirComp-based mechanisms Air-FedAvg and Dynamic. As shown, Air-FedGA outperforms Air-FedAvg and Dynamic in terms of convergence speed and accuracy. For instance, as shown in Fig. \ref{fig_LR_mnist}, Air-FedGA achieves an accuracy of 89.7\% after 5000s of training, while Air-FedAvg and Dynamic only reach 88.3\% and 82.5\%, respectively. Moreover, Air-FedGA attains a stable 80\% accuracy in 1077s, which is about 29.9\% and 71.6\% faster than Air-FedAvg (1536s) and Dynamic (3794s), respectively. The reason for the superior performance of Air-FedGA is that it adopts a group asynchronous updating mechanism that reduces the waiting time of workers, while Air-FedAvg suffers from long waiting time due to its synchronous updating. The loss and accuracy curves of jitter Dynamic more violently due to its selection of workers in each round, which introduces bias to the global model \cite{ma2024feduc}. In contrast, Air-FedGA groups workers considering the data distribution among workers, which makes inter-group data distribution as close to IID as possible. Therefore, Air-FedGA can handle Non-IID better and greatly reduce the jitter degree compared with Dynamic.

\subsubsection{Handling Edge Heterogeneity}

\begin{figure}[t]
\includegraphics[width=0.98\linewidth,height=3.6cm]{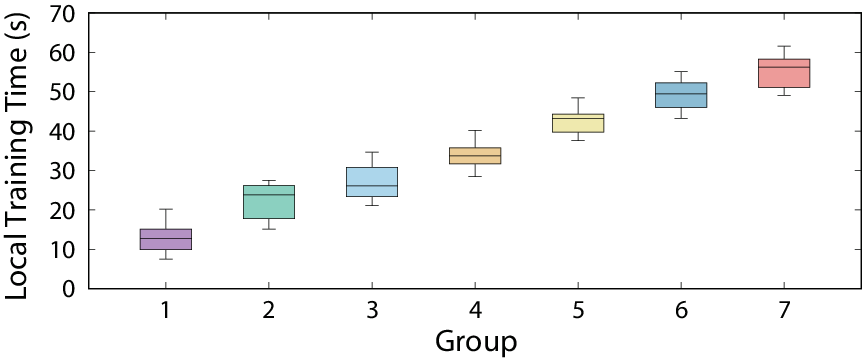}
\small{\caption{Grouping of workers with different local training time when $\xi=0.3$.}\label{fig_cluster}}
\vspace{-2mm}
\end{figure}
\begin{figure}[t]
\includegraphics[width=0.98\linewidth,height=3.6cm]{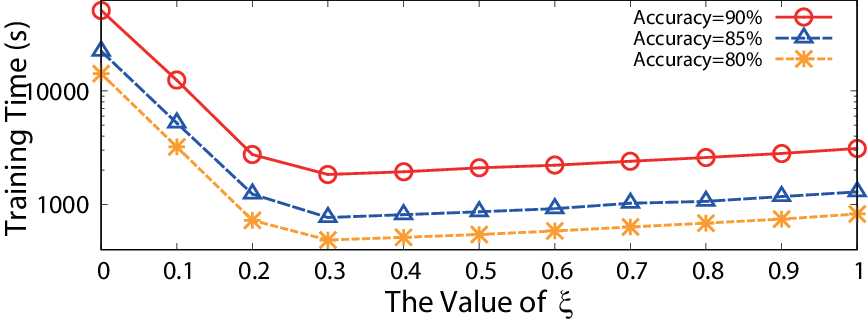}
\small{\caption{Training time under different values of $\xi$.}\label{fig_xi_time}}
\vspace{-2mm}
\end{figure}

To address edge heterogeneity, we intuitively tend to group workers with similar local training time together. Recall that the parameter $\xi$ in constraint \eqref{eq:14} quantifies the similarity of local training time of workers within each group. Fig. \ref{fig_cluster} presents a box plot illustrating the grouping of 100 workers with varying local training times when $\xi=0.3$. As shown in this figure, workers with comparable training time are generally clustered within the same group. For example, the local training times of the 100 workers range from 8.1s to 61.6s, while workers in Group 7 have local training times between 49.1s and 61.6s.

It is obvious that the number of workers in each group increases as the value of $\xi$ rises. To identify the optimal value of $\xi$, Fig. \ref{fig_xi_time} illustrates the training time required for CNN to achieve accuracy of 80\%, 85\%, and 90\% on the MNIST dataset, with $\xi$ ranging from 0 to 1. The results indicate that when $\xi = 0.3$, the training time to reach 80\%, 85\%, and 90\% accuracy is minimized, taking 485s, 765s and 1834s, respectively.
As $\xi$ approaches 0, required training time increases sharply. This is because smaller values of $\xi$ results in fewer workers per group, thereby limiting the benefits of reducing communication time in over-the-air aggregation. For example, when $\xi = 0$, each worker performs global updating in a fully asynchronous manner without over-the-air aggregation, and the training times increase significantly, reaching 14213s, 22426s and 51334s for 80\%, 85\% and 90\% accuracy, respectively. Conversely, as $\xi$ approaches 1, the training time also gradually increases. This is because the training duration for each group is determined by the worker with the longest local training time, exacerbating the straggler problem as the group size grows. At higher values of $\xi$, the number of workers in each group increases and ultimately decelerating FL. For example, when $\xi = 1$, the time required to reach 80\%, 85\% and 90\% accuracy is 823s, 1288s and 3110s, respectively.

\subsubsection{Handling Non-IID Data}

\begin{table}[t]\centering
\setlength{\belowcaptionskip}{0.05cm}
\caption{The impact of the grouping methods on EMD}\label{tbl:EMD}
\setlength{\tabcolsep}{4mm}{
\begin{tabular}{c|c|c|c} 
\hline
\makecell[c]{Methods} & Original & TiFL & Air-FedGA  \\ \hline
EMD & 1.8 & 0.69 & 0.21  \\ \hline
\end{tabular}}
\vspace{-2mm}
\end{table}

Table \ref{tbl:EMD} shows the average EMD $\overline{\Lambda}=\frac{1}{|\mathbf{V}|}\sum_{\mathcal{V}_j\in\mathbf{V}}\Lambda_j$ among groups after applying TiFL and Air-FedGA grouping methods.  As shown, the original average EMD is $\overline{\Lambda}=|\frac{1}{10}-1|+|\frac{1}{10}-0|\times9=1.8$, since each worker has data with the same label. After the grouping of TiFL, the EMD  is reduced to 0.69. However, by applying our proposed Air-FedGA, the EMD is further reduced to 0.21. This demonstrates that Air-FedGA can achieve a more balanced data distribution among groups, which is closer to IID.

\subsubsection{Energy Consumption for Over-the-air Aggregation}\label{subsubsec:convergencetime}

\begin{figure}[t]
\includegraphics[width=0.49\linewidth,height=3.6cm]{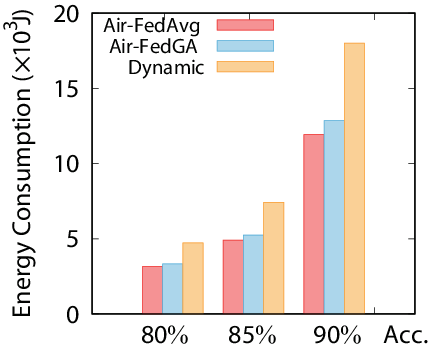}
\includegraphics[width=0.49\linewidth,height=3.6cm]{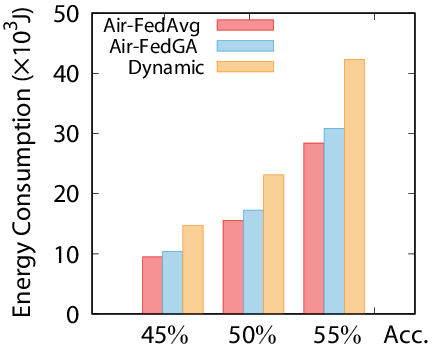}
\small{\caption{Energy Consumption vs. Accurary. \textit{Left}: CNN on MNIST; \textit{Right}: CNN on CIFAR-10.}\label{fig_energy_accuracy}}
\vspace{-2mm}
\end{figure}

Fig. \ref{fig_energy_accuracy} compares the model aggregation energy consumption of our Air-FedGA with two other AirComp-based mechanisms: Air-FedAvg and Dynamic. As shown, to achieve the same training accuracy, Air-FedGA consumes slightly more energy than Air-FedAvg, but less than Dynamic. The reason is that Air-FedGA performs asynchronous global updating among groups, which leads to more aggregations per worker on average than Air-FedAvg. Dynamic does not take into account the data distribution among workers when selecting a subset of workers for global updating, so it requires more global updating to converge. For instance, when training CNN on CIFAR-10, the model aggregation energy consumption of Air-FedAvg, Air-FedGA and Dynamic to reach 55\% accuracy is 28432J, 30856J and 42343J, respectively.

\subsubsection{Scalability}\label{subsubsec:convergencetime}

\begin{figure}[t]
\includegraphics[width=0.49\linewidth,height=3.6cm]{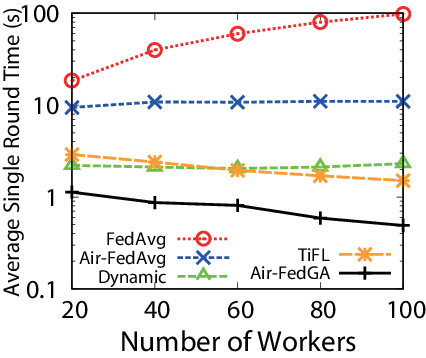}
\includegraphics[width=0.49\linewidth,height=3.6cm]{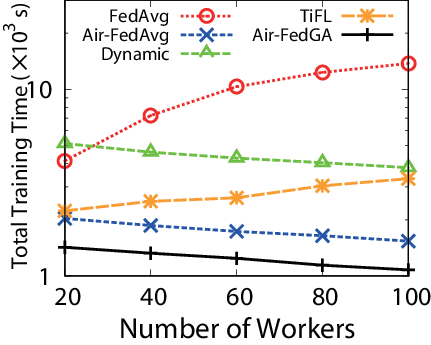}
\small{\caption{Training Time vs. Number of Workers. \textit{Left}: Single Round; \textit{Right}: Total.}\label{fig_time_M}}
\vspace{-2mm}
\end{figure}

Fig. \ref{fig_time_M} compares the average single round and total training time for CNN trained on MNIST of different methods, respectively, by varying the number $N$ of workers. Note that due to the significant difference in time in magnitude, we adopt logarithmic coordinates in this figure.
As shown in the left plot of Fig. \ref{fig_time_M}, the average single round training time for FedAvg grows with $N$, whereas Air-FedAvg and Dynamic remain relatively stable. This is because FedAvg requires each worker to upload its model to the server, which takes longer as $N$ increases, whereas Air-FedAvg and Dynamic adopt over-the-air aggregation, and the latter does not depend on $N$. On the other hand, the single round training time of Air-FedGA and TiFL decrease with $N$, since more workers lead to more groups, and the asynchronous participation of groups enables more frequent global updates. The right plot of Fig. \ref{fig_time_M} shows the total training time of different methods to achieve 80\% accuracy. It is observed that the training time of the methods without over-the-air aggregation increases with the increase of $N$, whereas that of the methods with over-the-air aggregation decreases with the increase of $N$. Consequently, the greater $N$, the more exponential performance advantage our Air-FedGA can show over other methods. For example, when $N=100$, the total training time of FedAvg, Dynamic, TiFL, Air-FedAvg and Air-FedGA is 13755s, 3799s, 3319s, 1536s and 1077s, respectively.

\section{Conclusion}\label{sec:conclusion}

In this paper, we have proposed an AirComp-based grouping asynchronous federated learning mechanism (Air-FedGA) to address the challenges of communication resource constraint, heterogeneity and data Non-IID at network edge. The proposed mechanism allows FL to accelerate the model training by over-the-air aggregation, while relaxing the synchronization requirement of this aggregation technology. We have analyzed the convergence of Air-FedGA and formulated a training time minimization problem, which jointly optimizes the power scaling factors at edge devices, the denoising factors at the parameter server, and the worker grouping strategy. We have provided the power control and worker grouping algorithms to solve this problem. Extensive simulations demonstrate that the proposed solutions significantly accelerate FL compared with the state-of-the-art solutions, while effectively handling heterogeneity, Non-IID data, and ensuring scalability.

\section*{Acknowledgement}
The corresponding author of this paper is Junlong Zhou. This work was supported in part by the National Science Foundation of China (NSFC) under Grants 62402537, 62172224 and 92367104, in part by the Natural Science Foundation of Jiangsu Province under Grant BK20220138, in part by the Jiangsu Province Excellent Postdoctoral Program under Grant JB23085, and in part by the Fundamental Research Funds for the Central Universities under Grant 30922010318.
\newpage

\bibliographystyle{IEEEtran}
\bibliography{contents/refs}

\appendices

\end{document}